\documentclass[review,12pt]{elsarticle}

\usepackage{lineno,hyperref}

\usepackage[margin=1.in]{geometry}
\usepackage{setspace}

\usepackage{hyperref}
\hypersetup{
    colorlinks = true,
     }
\usepackage{amsmath}

\biboptions{sort&compress}
\usepackage{times}

\begin{document}

\begin{frontmatter}

\title{A finite element model for seismic response analysis of free-standing rocking columns with vertical dampers}
\author{Mehrdad Aghagholizadeh\corref{cor1}}
\cortext[cor1]{Corresponding author}
\ead{mehrdadag@smu.edu}
\address{Lecturer, Department of Civil and Environmental Engineering, Southern Methodist University\\P.O. Box 750340 Dallas, TX 75275-0340}


\begin{abstract}
This paper investigates finite-element modeling of a vertically damped free-standing rocking column. The paper first derives the nonlinear equation of motion for the coupled system and then compares the analytical solution with finite-element model. Finite-element model is being produced using open source framework named OpenSees. The rocking surface is modeled using zero-length fiber cross-section element and the dampers are modeled with two node link elements. In order to simulate energy dissipation during the rocking motion Hilber-Hughes-Taylor numerical dissipative time step integration is being adopted. The paper also compares two types of hysteretic and viscous damping devices and it shows that the viscous damping behavior is favorable when it is used along with a rocking block. The results of analytical model of a rigid block with viscous dampers in MATLAB is then compared with OpenSees model and the paper concludes that the finite-element model compares satisfactory with the analytical model.
\end{abstract}


\end{frontmatter}


\section{Introduction}

Early study of seismic response of free-standing blocks has been presented in works of Milne in 1885 \cite{milne1885seismic}. Milne's work was a static analysis of rocking objects and he concluded that if the ground acceleration exceeds $g\cdot\frac{width}{height}$, the rocking column will overturn. More than 40 years after Milne's work, Kirkpatrick \cite{kirkpatrick1927seismic}, Ikegami and Kishinouye \cite{ikegami194910,ikegami1950acceleration} and Muto et al. \cite{muto1960study} published remarkable works on seismic response of rocking columns. Kirkpatrick's work \cite{kirkpatrick1927seismic} introduced effects of two important quantities of column size and the period of the excitation to be considered in the analysis of the rocking objects. In 1963, Housner \cite{housner1963behavior} showed that tall, slender, free-standing columns, while they can easily uplift even when subjected to a moderate ground acceleration (uplift initiates when $\ddot{u}_g>g\times(base/height)$); they exhibit remarkable seismic stability because of a size-frequency scale effect. His work shows that there is a safety margin between uplifting, overturning and that as the size of the free-standing column increases or the frequency of the excitation pulse increases, this safety margin increases appreciably to the extent that large free-standing columns enjoy ample seismic stability. Zhang and Makris \cite{zhang2001rocking} and Makris and Konstantinidis \cite{makris2003rocking} studied the response of rocking blocks under pulse-like records. Zhang and Makris \cite{zhang2001rocking} showed that there are two modes of overturning for rocking objects,  the first one is by exhibiting one or more impacts; and the second mode is without exhibiting any impact. This work concludes that because of the nonlinear nature of the problem, in association with the presence of the safe region, complicates the task of estimating the peak ground acceleration by only examining the geometry of freestanding objects that either overturned or survived a ground shaking.

\noindent In the late 1960s in New Zealand the concept of allowing a tall, slender structure to uplift and rock was first implemented with the design of South Rangitikei Rail Bridge \cite{beck1972seismic,beck1973seismic,bookKelly,skinner1993introduction}. The 72~m tall bridge piers are designed to uplift and rock around their pivoting points \cite{kelly1972mechanisms,skinner1974hysteretic}. The rocking response of each pier is damped with a pair of torsionally yielding steel-beam dampers \cite{skinner1980hysteretic}.

\noindent Following Housner (1963), number of publications studied the rocking response of free-standing blocks and columns \cite{Aslam1980,ishiyama1982motions,spanos1984rocking,hogan1990many,tso1989steady} and recently the dynamics of rocking objects has received increasing attention \cite{makris2002uplifting,roh2010modeling,makris2012sizing,
dimitrakopoulos2012overturning,dejong2014dynamically,kalliontzis2016improved,
makris2016size,aghagholizadeh2017seismic,kalliontzis2017improved,makris2017Bearthquake,makris2017dynamics,makris2017earthquake,agg7,chatzis2018energy,
vassiliou2018seismic,sun2018practical,gioiella2018modal,makris2019effect,nazari2019seismic}.

\noindent Vassiliou et al. \cite{vas2014dynamic,vas2017finite} studied the finite element model of the solitary rocking blocks and frames. In their study, to define the rocking surface, they assigned a fiber element with zero-length and a material with a relatively high (almost rigid) modulus of elasticity in compression but no tension strength (similar approach is also adopted in this study). Dimitrakopoulos and DeJong \cite{dimitrakopoulos2012overturning} and Makris and Aghagholizadeh \cite{makris2019effect} studied dynamics of a rocking column when equipped with dampers. This studies, investigate the analytical model of the rocking block-damper.

\noindent Considering the growing interest for response modification using rocking isolation, the main objective of this study is to help researchers and professional engineers to be able to model the rocking columns, bridge piers or walls when they are equipped with dampers, using the currently available tools. It is worth mentioning that this paper studies dynamics of a rigid rocking blocks, therefore, effects of column flexibility needs further investigation. This study first derives nonlinear equation of motion for a rocking column with viscous dampers. Then comparison of force-displacement relation of hysteretic dampers with viscous dampers are discussed and viscous damping is adopted for the analysis. Lastly, a numerical model using OpenSees framework is developed to calculated in-plane response of the system. Response of the finite-element model under different ground motions is compared with the analytical solutions from MATLAB. 

\section{Geometric Parameters of a Rocking Column with Vertical Dampers}
\begin{figure}
\centering
\includegraphics[scale=0.65]{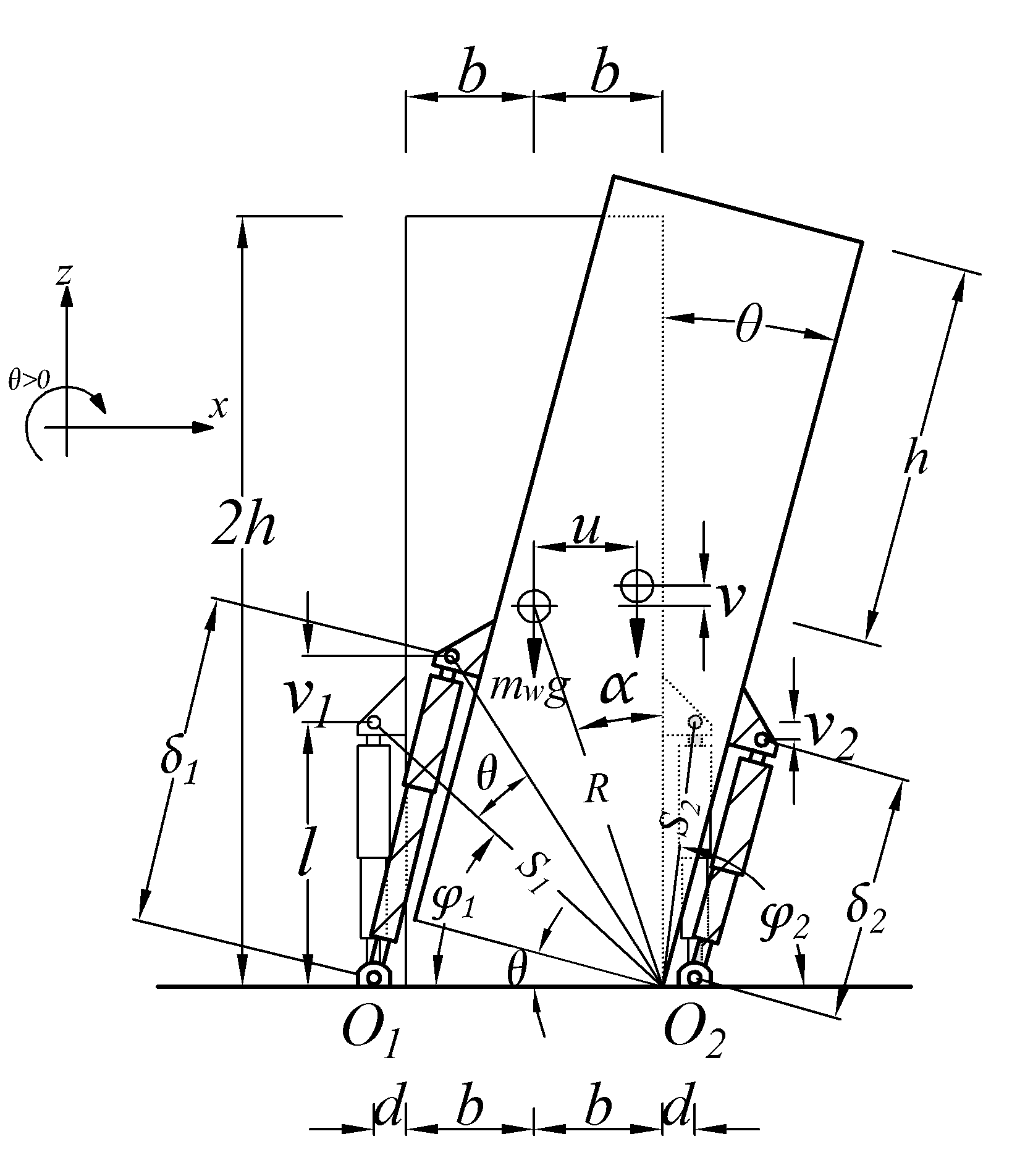}
\caption{Geometric properties of a rocking column with additional energy dissipators.}
\label{fig:1}
\end{figure}

Figure (\ref{fig:1}) shows geometric properties and parameters of a solitary rigid-rocking-column when it is coupled with vertical dampers. Considering the geometry of the problem, this study investigates the dynamic response of a the rocking column with width $2b$, height $2h$, size $R=\sqrt{b^2+h^2}$, slenderness, $\tan \alpha=b/h$, mass $m$, and moment of inertia about the pivoting (stepping) points $O$ and $O'$, $I=\frac{4}{3}mR^2$. Vertical energy dissipation devices are mounted to the rocking column at a distance, $d$, from the pivoting points of the column as shown in Figure (\ref{fig:1}). The upward displacement of the damper located across from the pivoting point, and the downward displacement; $v_2$ of the damper located at the same side as the pivoting corner of the column are
\begin{equation} \label{eq:v1}
v_1=S_1\big[\sin(\phi_1 \pm \theta)-sin\phi_1\big] 
\end{equation}
\begin{equation} \label{eq:v2}
v_2=S_2\big[sin\phi_2-\sin(\phi_2 \mp \theta)\big] 
\end{equation}
\noindent where $S_1=\sqrt{(2b+d)^2+l^2}$, $S_2=\sqrt{d^2+l^2}$, $sin\phi_1=l/S_1$ and $\sin\phi_2=l/S_2$.

\noindent In all equations provided in this study, when the double sign (e.g. $\pm$) is used, the top sign is for $\theta>0$ and the bottom sign is for $\theta<0$.

\noindent The elongation of the damper located across from the pivoting point is $e_1$ and the contraction of the damper at the pivoting point side is $e_2$, these parameters are calculated as follows:
\begin{equation} \label{eq:e1}
{e_1}(t) = {S_1}\left[ {\sqrt {1 + {{\cos }^2}{\varphi _1} - 2\cos {\varphi _1}\cos ({\varphi _1} \pm \theta )}  - \sin {\varphi _1}} \right]
\end{equation}
\begin{equation} \label{eq:e2}
{e_2}(t) = {S_2}\left[ {\sin {\varphi _2} - \sqrt {1 + {{\cos }^2}{\varphi _2} - 2\cos {\varphi _2}\cos ({\varphi _2} \mp \theta )} } \right]
\end{equation}
and the time derivatives of the equations (\ref{eq:e1}) and (\ref{eq:e2}) are:
\begin{equation} \label{eq:e1dot}
{\dot e_1}(t) = \frac{{  {S_1}\cos {\varphi _1}\,\dot \theta \sin ({\varphi _1} \pm \theta )}}{{\sqrt {1 + {{\cos }^2}{\varphi _1} - 2\cos {\varphi _1}\cos ({\varphi _1} \pm \theta )} }}
\end{equation}
\begin{equation} \label{eq:edot2}
{\dot e_2}(t) = \frac{{ {S_2}\cos {\varphi _2}\dot \theta \sin ({\varphi _2} \mp \theta )}}{{\sqrt {1 + {{\cos }^2}{\varphi _2} - 2\cos {\varphi _2}\cos ({\varphi _2} \mp \theta )} }}
\end{equation}

\section{Equation of Motion of a Rigid Rocking-Column with Dampers}
In order to find governing equation of motion for the coupled system, the responses are calculated for two cases of positive $\theta(t)>0$ and negative $\theta<0$ rotation.

\noindent For positive rotations ($\theta>0$), dynamic equilibrium of the rocking column with mass, $m$, equipped with vertical dampers installed on each of its sides as shown in Figure (\ref{fig:1}) gives (in all following equations subscript 1 refers to the damper across the pivoting point and 2 refers to damper next to the pivoting point.)

\begin{equation} \label{eq:EOMg1}
I\ddot \theta  =  - mgR\sin (\alpha  - \theta ) - m{\ddot u_g}R\cos (\alpha  - \theta ) - {F_{{d_1}}}{r_1} - {F_{{d_2}}}{r_2}
\end{equation}
\noindent in which $F_{d_1}$and $F_{d_2}$ are the damping forces from the dampers and $r_1$ and $r_2$ are the respective moment arms of the damping forces about the pivoting points

\begin{equation} \label{eq:r1}
{r_1} = {S_1}\cos {\varphi _1}\frac{{\sin ({\varphi _1} + \theta )}}{{\sqrt {1 + {{\cos }^2}{\varphi _1} - 2\cos {\varphi _1}\cos ({\varphi _1} + \theta )} }}
\end{equation}
\noindent and

\begin{equation} \label{eq:r2}
{r_2} = {S_2}\cos {\varphi _2}\frac{{\sin ({\varphi _2} - \theta )}}{{\sqrt {1 + {{\cos }^2}{\varphi _2} - 2\cos {\varphi _2}\cos ({\varphi _2} - \theta )} }}
\end{equation}
\noindent Hence, the equation of motion can be written as:

\begin{equation} \label{eq:EOMg2}
\ddot \theta  =  - {p^2}\left[ {\sin (\alpha  - \theta ) + \frac{{{{\ddot u}_g}}}{g}\cos (\alpha  - \theta ) + \frac{{{F_{{d_1}}}}}{{mg}}\frac{{{r_1}}}{R} + \frac{{{F_{{d_2}}}}}{{mg}}\frac{{{r_2}}}{R}} \right]
\end{equation}
\noindent in which, $p=\sqrt{mgR/I}$, is the frequency parameter of the column and for a rectangular column $I=4/3mR^2$, which leads to $p=\sqrt{3g/4R}$ \cite{makris2014half,makrisc1999rocking}.

\noindent For negative rotation ($\theta<0$), one can follow the same reasoning and the equation of motion is

\begin{equation} \label{eq:EOMg1neg}
\ddot \theta  =  - {p^2}\left[ { - \sin (\alpha  + \theta ) + \frac{{{{\ddot u}_g}}}{g}\cos (\alpha  + \theta ) + \frac{{{F_{{d_1}}}}}{{mg}}\frac{{{r_1}}}{R} + \frac{{{F_{{d_2}}}}}{{mg}}\frac{{{r_2}}}{R}} \right]
\end{equation}

\noindent and equations(\ref{eq:EOMg2}) and (\ref{eq:EOMg1neg}) can be expressed in the compact form using signum function, $sgn(\theta)$:
\begin{equation} \label{17}
\ddot \theta  =  - {p^2}\left\{ {\sin (\alpha {\mathop{\rm sgn}} \theta  - \theta ) + \frac{{{{\ddot u}_g}}}{g}\cos (\alpha {\mathop{\rm sgn}} \theta  - \theta ) + \frac{{{F_{{d_1}}}}}{{mg}}\frac{{{r_1}}}{R} + \frac{{{F_{{d_2}}}}}{{mg}}\frac{{{r_2}}}{R}} \right\}
\end{equation}

\noindent During the oscillatory rocking motion of a free-standing rigid column (no dampers) energy is lost only during impact, when the angle of rotation reverses. When the angle of rotation reverses, it is assumed that the rotation continues smoothly from points $O_2$ to $O_1$ (Figure \ref{fig:1}) and that the impact force is concentrated at the new pivot point $O_1$. With this idealization, the impact force applies no moment about $O_1$; hence, the angular momentum is conserved. Conservation of angular momentum about point $O_1$ just before the impact and right after the impact gives that the angular velocity after the impact=$\dot{\theta}_a$ is only $\eta$ times the angular velocity before the impact=$\dot{\theta}_b$; where $\eta$ depends solely on geometry

\begin{equation} \label{eq:rest}
\eta  = \frac{{{{\dot \theta }_a}}}{{{{\dot \theta }_b}}} = 1 - \frac{3}{2}{\sin ^2}\alpha
\end{equation}

\noindent The restitution factor calculated by equation (\ref{eq:rest}) is only the minimum energy dissipation required during impact for a column with slenderness $\alpha$ to engage in rocking motion (not to jump) \cite{makris2014half}. For the case of slender rocking blocks, because of inelastic behavior at the impact, the true value of coefficient of restitution will be less than the one computed by the equation.  In the recent research, Kalliontzis et al. \cite{kalliontzis2016improved} introduced an improved coefficient of restitution estimation that can predict this value more accurately. In this study, the analysis is based on the equation (\ref{eq:rest}).

\section{Viscous Dampers}
In a rocking rigid block, during the oscillatory motion, one source of the energy dissipation is the energy that is lost during the impact when the angle of rotation reverses \cite{makris2014half}. In order to increase energy dissipation of a rocking block during seismic excitation, different types of damping systems can be added to the rocking column \cite{makris2001rocking,ajrab2004rocking,marriott2008dynamic,toranzo2009shake,roh2010modeling,dimitrakopoulos2012overturning}, this study implements viscous dampers. 

\begin{figure}[t]
\centering
\includegraphics[scale=0.85]{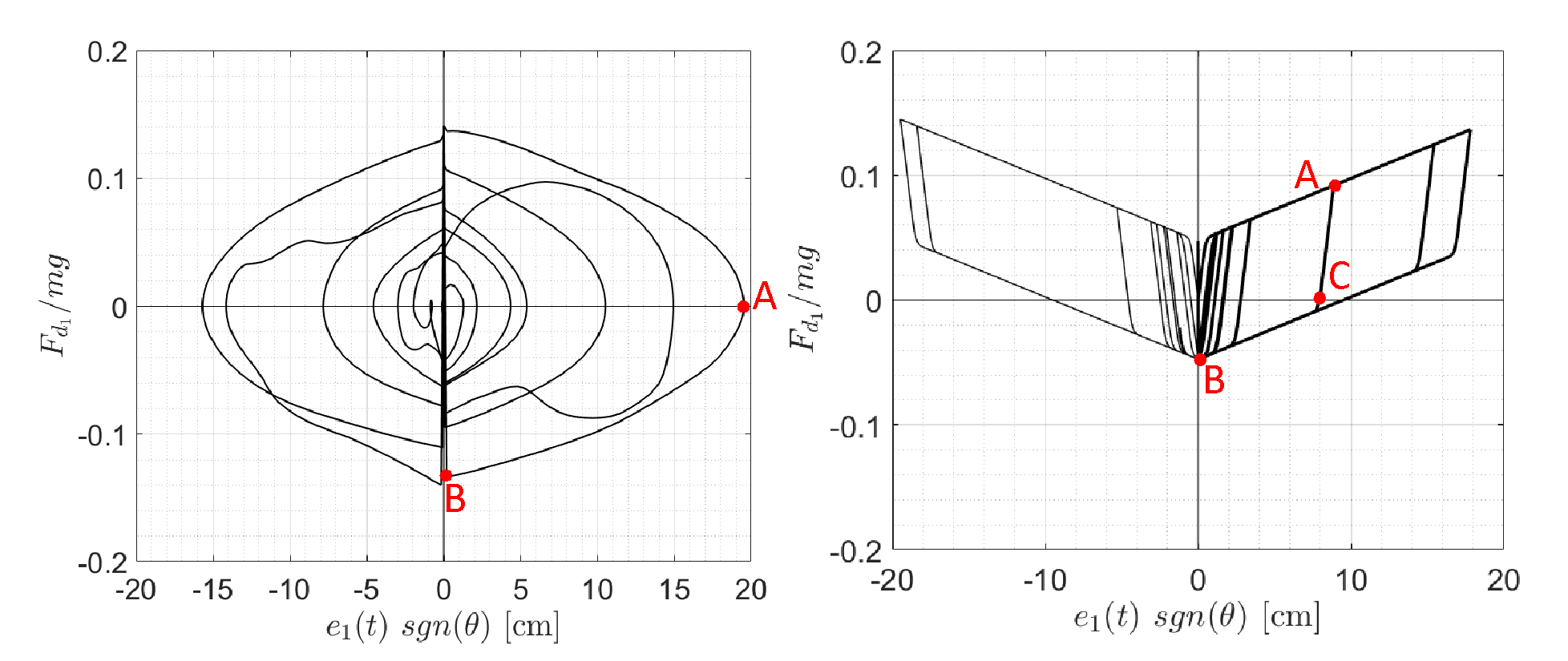}
\caption{Force-Displacement relation comparison of a viscous damper (left) with a hysteretic damper (right).}
\label{fig:Visc_Hyst}
\end{figure}

\noindent Figure \ref{fig:Visc_Hyst} shows comparison of force-displacement relationship of a viscous damper when compared with a hysteretic damper. When a viscous damper is used at the corner of a rocking block, at the maximum displacement, the rocking block has zero velocity; therefore, the damper has zero force (point A in the figure \ref{fig:Visc_Hyst} (left)). When the rocking object is returning to its pivoting point (from point A to B in the figure \ref{fig:Visc_Hyst} (left)), the viscous damper has a negative force which will create an opposite moment compared to the moment created by the rocking block's self-weight (equation \ref{17}). However, when a rocking block is equipped with a hysteretic damper, when the rocking object is returning from its maximum displacement, the damper first have a positive force (from A to C in the figure \ref{fig:Visc_Hyst} (right)) and then it starts creating a negative force (from C to B in the figure \ref{fig:Visc_Hyst} (right)). This phenomenon can create a harder impact and in some cases can increase the response of damped rocking block compared to undamped one. Because of the mentioned difference between the force-displacement behaviors of these two types of damping system, the viscous damper is used in this study.

\noindent The dissipation devices attached to the rocking column (Figure \ref{fig:1}) can be either linear or nonlinear fluid dampers \cite{cons1998pa,symans1999semi,so1999pass,symans2008energy}. For the general case of viscous dampers, the constitutive law of the nonlinear viscous damper is \cite{cons1998pa,so1999pass}:

\begin{equation} \label{eq:FdVISC}
{F_d} = {C_q}\,{\left| {\dot e(t)} \right|^q}{\mathop{\rm sgn}} \left[ {\dot e(t)} \right]
\end{equation}

\noindent where $0<q<1$ is the exponent of the damper, $C_q$ is the damping constant with units: $[m] [L]^{(1-q)} [T]^{(q-2)}$,  and $sgn[~]$ is the signum function. When $q=1$, equation (\ref{eq:FdVISC}) reduces to a linear viscous law: $F_d=c_1 \dot{e}(t)$.

\section{Dynamics of a Solitary Column with Supplemental Linear Viscous Dampers at Pivoting Points}

For the special case, when zero-length ($l=\phi_1=\phi_2=0$) linear viscous dampers are used at the pivoting points ($d=0,S_1=2b,S_2=0$) therefore equations (\ref{eq:e1}) and (\ref{eq:r1}) simplify to

\begin{equation} \label{eq:e_1andr_1}
{e_1} = 2\sqrt 2 b\sqrt {1 - \cos \theta }  \quad \quad and \quad  \quad {r_1} = \sqrt 2 \frac{{b\sin \theta }}{{\sqrt {1 - \cos \theta } }}
\end{equation}

\noindent while the time derivative of the stroke is given by

\begin{equation} \label{eq:e_1dot2}
{\dot e_1}(t) = {\sqrt 2 \,b\dot \theta}{\sqrt {1 + \cos \theta } }
\end{equation}

\noindent which for small rotations simplifies to $\dot{e}_1(t)=2b\dot{\theta}$. The full nonlinear equation of the stepping pier with zero-length viscous (linear or nonlinear) dampers at its pivoting points is
\begin{equation} \label{eq:ViscEOM}
\ddot \theta  =  - {p^2}\left\{ {\sin (\alpha  - \theta ) + \frac{{{{\ddot u}_g}}}{g}\cos (\alpha  - \theta ) + \frac{{\sqrt 2 \,\sin \alpha }}{{mg}}\sqrt {1 + \cos \theta } \,{C_q}{{\left| {\sqrt 2 \,b\dot \theta \,\sqrt {1 + \cos \theta } } \right|}^q}{\mathop{\rm sgn}} [\dot \theta ]} \right\}
\end{equation}
\noindent and for the case of zero-length linear ($q=1$) viscous damper, equation (\ref{eq:ViscEOM}) becomes
\begin{equation} \label{eq:LinViscEOM}
\ddot \theta  =  - {p^2}\left\{ {\sin (\alpha  - \theta ) + \frac{{{{\ddot u}_g}}}{g}\cos (\alpha  - \theta ) + \frac{3C_1 \sin^2\alpha}{2mp^2}}(1+\cos\theta)\dot{\theta} \right\}
\end{equation}
\noindent which confirms with equation presented in \cite{dimitrakopoulos2012overturning,makris2019effect}. 

\section{Finite Element Model for Rocking Column with Supplemental Viscous Dampers}
\subsection{Rocking Surface Model}
\begin{figure}[t]
\centering
\includegraphics[scale=0.35]{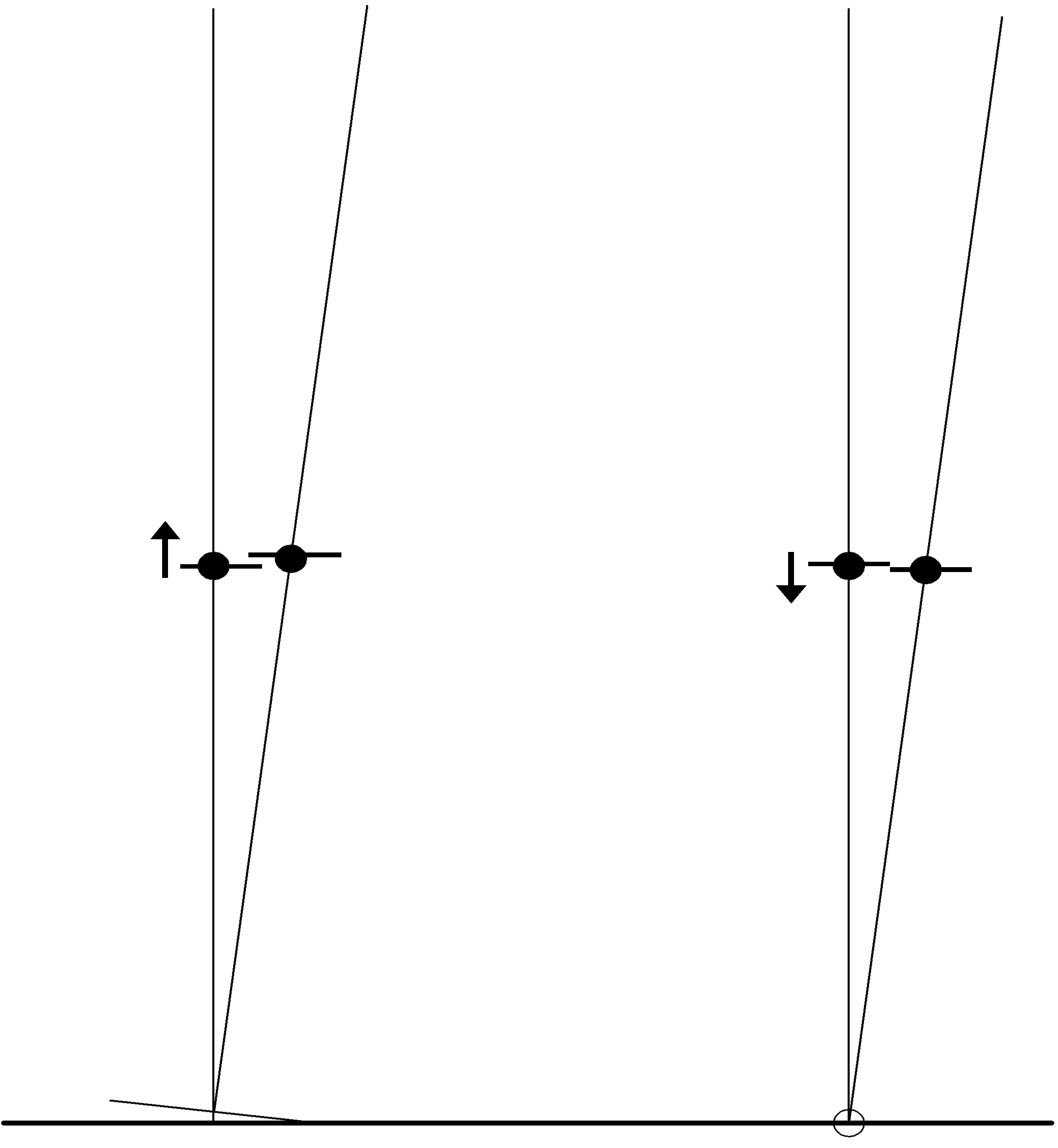}
\caption{Comparison of the vertical displacement of a point for the zero-length fiber element at the rocking surface (left) with the spring model for the rocking surface (right).}
\label{fig:ENT_vs_spring}
\end{figure}

There are two common approaches in order to model rocking column \cite{vas2014dynamic,vas2017finite} in OpenSees \cite{opensees,mazz2006opensees,fmk2011opensees}. One is defining the rocking surface as a rotational spring \cite{vas2014dynamic} and the other is defining a surface between bottom of the column and the ground using a zero-length element with an Elastic-No-Tension (ENT) material \cite{vas2017finite}. In this study, the latter is used. The advantage of zero-length element compared to rotational spring model is that the zero-length element captures vertical displacements of a point on a rocking block accurately, which is required when using vertical tendons or dampers. This is because the damping force or the force in a restraining tendon is a function of the tendon elongation or the relative velocity in a damper. In a rotational spring model, the vertical displacement is not calculated correctly. Figure (\ref{fig:ENT_vs_spring}) shows comparison of a point on a rocking block in a zero-length fiber element compared with the rotational spring model.

\noindent To model the rocking surface in OpenSees, zero-length fiber section element is placed between the bottom of the column and the ground surface. The cross-section of the zero-length rocking surface is defined, using of a fiber section with nonlinear, elastic-no-tension (ENT) material.

\noindent In order to consider energy dissipation of the rocking motion during the impact, similar to \cite{vas2017finite}, dissipative time-stepping integration procedure of Hilber-Hughes-Taylor \cite{HHT1977} (HHT) is being used. HHT damping effect is function of dissipation factor $a'_d$ and the time of the integration. For this study the HHT time step is selected as $10^{-4} s$ and dissipation factor $a'_d$ is selected as $-1/3$ \cite{vas2017finite}.

\noindent The column is defined as an elastic beam column element with relatively large modulus of elasticity to represent a rigid rocking column (E=$10^{11}$ kPa).
\begin{figure}[t]
\centering
\includegraphics[scale=0.5]{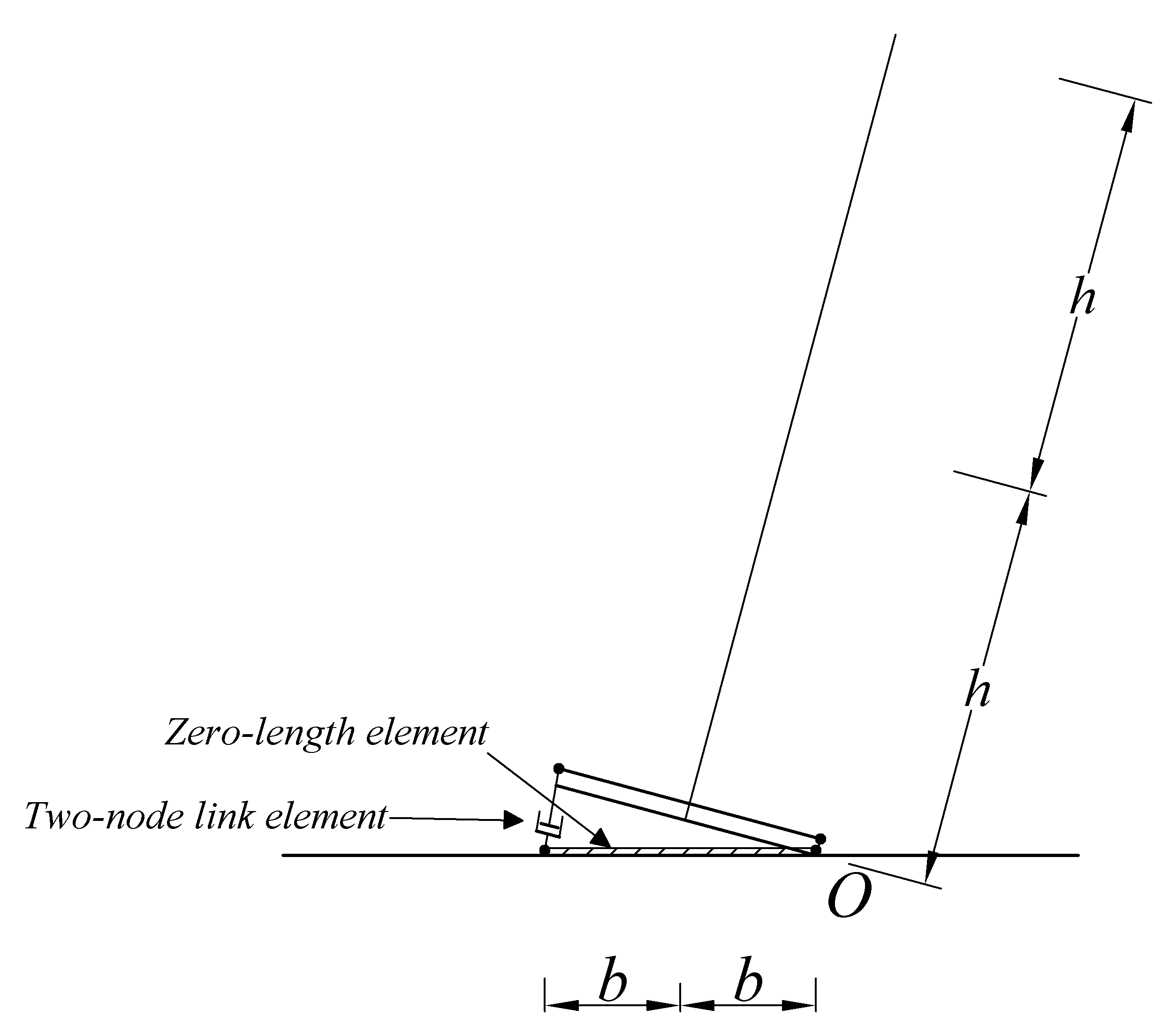}
\caption{Schematic for Finite element model of a rocking column.}
\label{fig:2}
\end{figure}

\begin{table}[b]
\centering
\caption{Dimensions and Material Properties for The Numerical Model}
\label{tab:my-table}
\begin{tabular}{lllll}\hline
\multicolumn{3}{|l|}	{\textbf{Parameter}}      & \multicolumn{2}{l|}	{\textbf{Value}} \\ \hline

\multicolumn{3}{|l|}	{Width}      & \multicolumn{2}{l|}	{2 m} \\ \hline
\multicolumn{3}{|l|}	{Height}      & \multicolumn{2}{l|}	{12 m} \\ \hline

\multicolumn{3}{|l|}	{$\tan(\alpha)$}      & \multicolumn{2}{l|}	{$1/6$} \\ \hline
\multicolumn{3}{|l|}	{Modulus of Elasticity of the Rigid Column, $E$}      & \multicolumn{2}{l|}	{$10^{11}$ kPa } \\ \hline
\multicolumn{3}{|l|}	{Modulus of Elasticity of the Fiber for Rocking Surface (ENT), $E_s$}      & \multicolumn{2}{l|}	{$30 \times 10^{9}$ kPa } \\ \hline

\multicolumn{3}{|l|}	{Viscous Damping Constant, $C$}      & \multicolumn{2}{l|}	{$3461.5$ Mg/sec } \\ \hline

\multicolumn{3}{|l|}	{HHT time step \quad~~}      & \multicolumn{2}{l|}	{$10^{-4} s$} \\ \hline
\multicolumn{3}{|l|}	{Dissipation factor $a'_d$}      & \multicolumn{2}{l|}	{$-1/3$ } \\ \hline
\end{tabular}
\end{table}

\subsection{Undamped Model}
\begin{figure}[b!]
\centering
\includegraphics[scale=1]{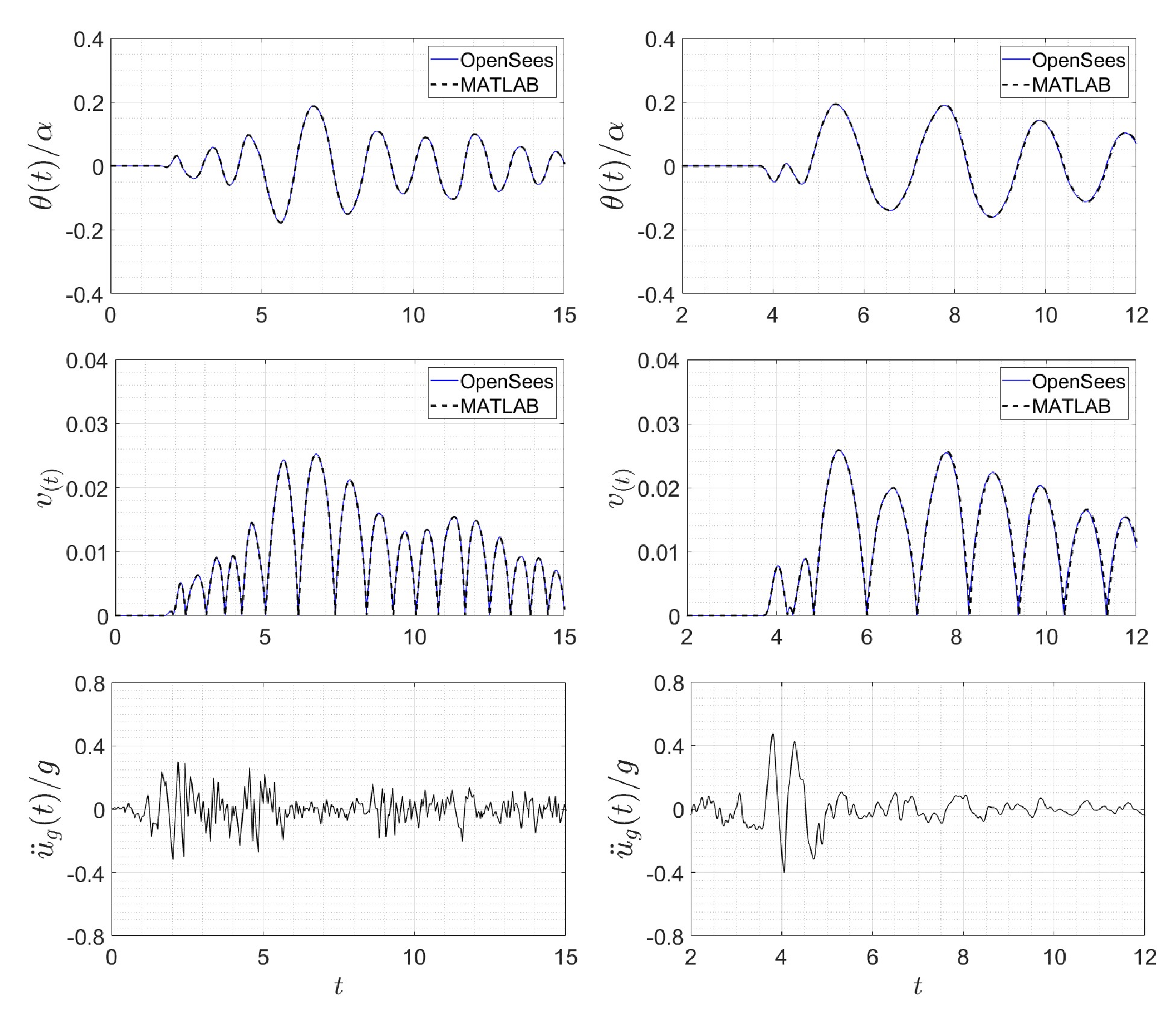}
\caption{Time history response comparison of an undamped rocking column modeled in MATLAB with OpenSees. The last row plots are time history records of El Centro ground motion recorded during the 1940 Imperial Valley, California (left) and CO2/064 ground motion recorded during the 1966 Parkfield, California (right) earthquakes.}
\label{fig:3}
\end{figure}

\noindent Figure (\ref{fig:3}) plots solution comparison of analytical model using equation (\ref{eq:LinViscEOM}) in MATLAB \cite{MATLAB2019} with OpenSees analysis of an undamped solitary rocking column when subjected to El Centro ground motion recorded during the 1940 Imperial Valley, California (left) and CO2/064 ground motion recorded during the 1966 Parkfield, California (right) earthquakes. The column has height of $12m$ and width of $2m$ with slenderness $\tan\alpha=1/6$. The top plots are comparison of the column normalized rotation with respect to its slenderness ($\theta(t)/\alpha$) in which it can be seen that the finite element model represents the rocking motion perfectly. The plots in the middle are comparison of top node vertical displacement from MATLAB and OpenSees. This comparison verifies the accuracy of the finite-element model in terms of column rotation, vertical displacement and energy dissipation of the impact.

\begin{figure}[t!]
\centering
\includegraphics[scale=.85]{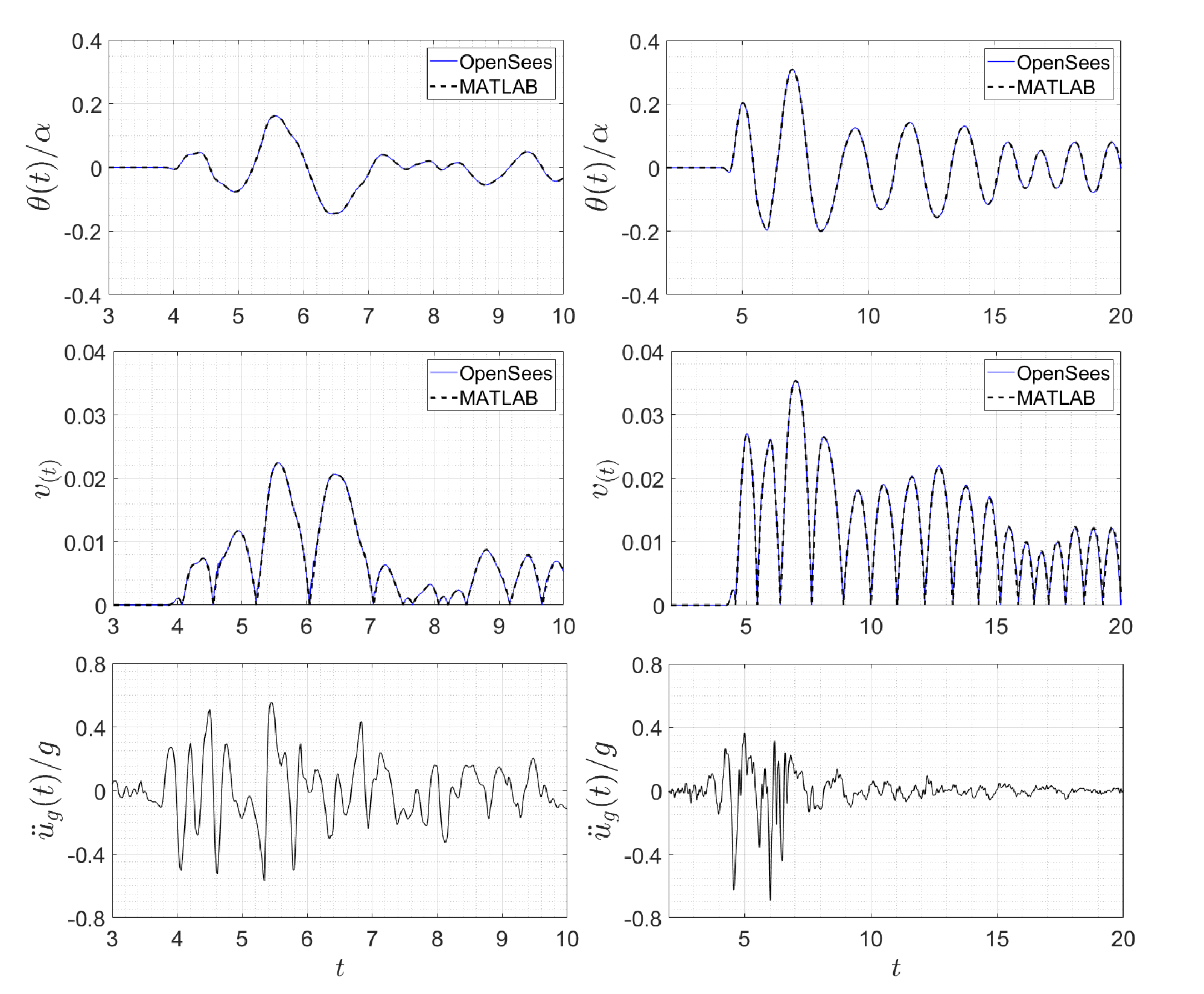}
\caption{Time history response comparison of an undamped rocking column modeled in MATLAB with OpenSees. The last row plots are time history records of Newhall/360 ground motion recorded during the 1994 Northridge, California (left) and Takarazuka/000 ground motion recorded during the 1995 Kobe, Japan (right) earthquakes.}
\label{fig:4}
\end{figure}

\noindent Figure (\ref{fig:4}) plots similar analysis with Figure (\ref{fig:3}) this time when column is subjected to Newhall/360 ground motion recorded during the 1994 Northridge, California (left) and Takarazuka/000 ground motion recorded during the 1995 Kobe, Japan (right) earthquakes. This figure is also confirms the accuracy of the finite element model compared to analytically acquired results.

\noindent In order to further examine the accuracy of the finite element model for the aforementioned undamped rocking column the model is also subjected to smooth mathematical pulses that can approximate the coherent, distinguishable pulses of recorded strong ground motions. Such mathematical acceleration pulses can be either simple rectangular pulses \cite{Hall1995Near,Alavi2001Eff}, trigonometrix pulses \cite{Vel1965Def,Mak1997,Mak2000} or more sophisticated wavelet functions \cite{Mav2003,Vas2011Est}.

\noindent In this study symmetric Ricker wavelet function (which is the second derivative of the Gaussian function, $e^{-t^2/2}$) and antisymmetric Ricker wavelet function (which is the third derivative of the Gaussian function) is used to simulate pulse-like records \cite{Rick1943,Rick1944}.

\begin{equation}\label{eq:Ricker}
\psi(t)=a_p\Big(1-\frac{2\pi^2t^2}{T_p^2}\Big)e^{-\frac{1}{2}\frac{2\pi^2t^2}{T_p^2}}
\end{equation}
\begin{figure}[b!]
\centering
\includegraphics[scale=.85]{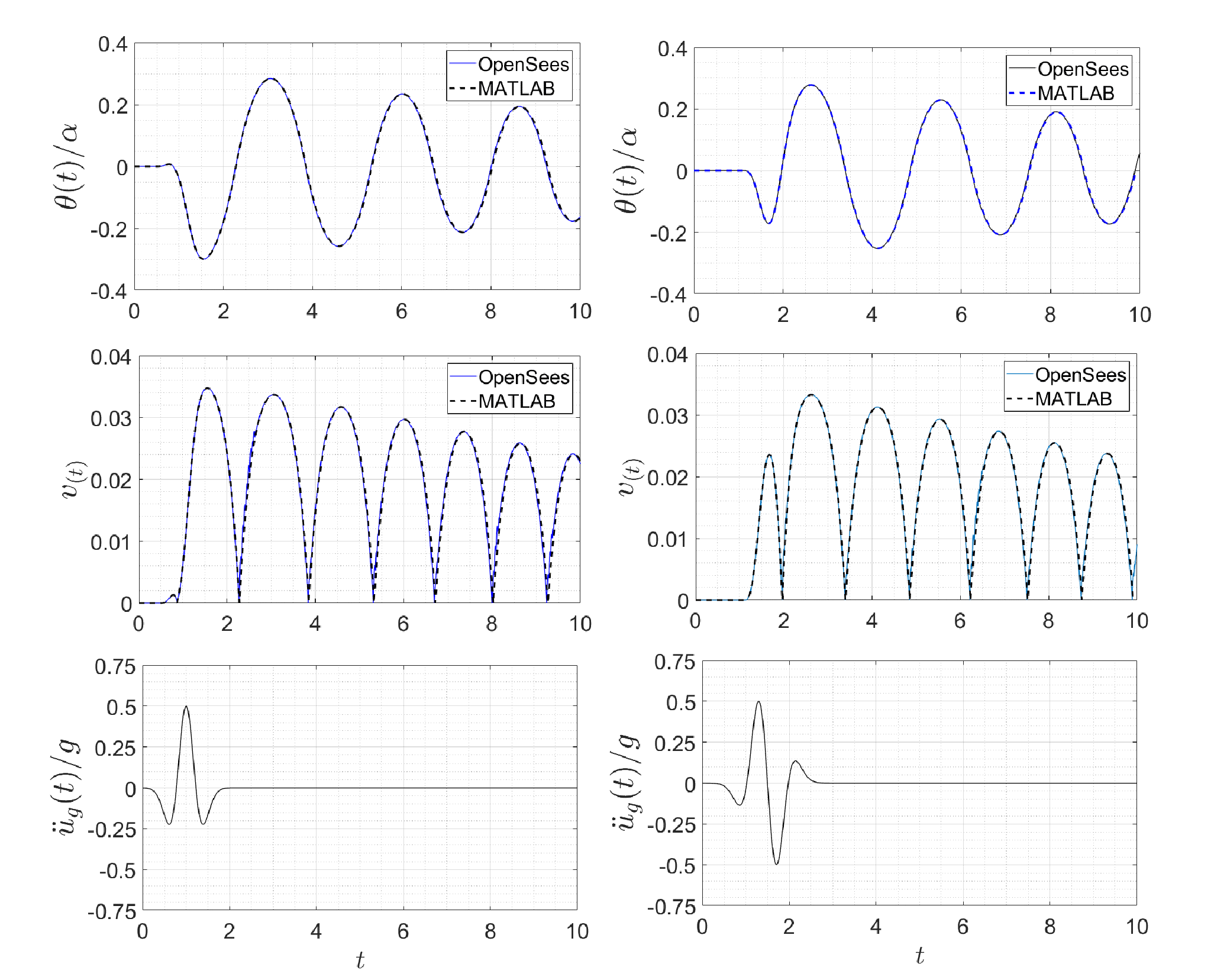}
\caption{Time history response comparison of an undamped rocking column modeled in MATLAB with OpenSees. The last row plots are time history records of subjected to a symmetric Ricker wavelet with acceleration amplitude ($a_p=0.5g$) (left) and antisymmetric Ricker wavelet with acceleration amplitude ($a_p=0.5g$) (right).}
\label{fig:5}
\end{figure}
\noindent Equation (\ref{eq:Ricker}) is the function for symmetric Ricker wavelet. The value of $T_p=\frac{2\pi}{\omega_p}=\pi\sqrt{2}s$ is the period that maximizes the Fourier spectrum of the symmetric Ricker wavelet; whereas, the time scale $s$, is the time from the peak acceleration of the wavelet to its first zero-crossing.

\noindent and the antisymmetric Ricker wavelet function is defined as:
\begin{equation}\label{eq:antiSymRicker}
\psi(t)=\frac{a_p}{\beta}\Big(\frac{4\pi^2}{3T_p^2}-3\Big)\frac{2\pi t}{\sqrt{3}T_p}e^{-\frac{1}{2}\frac{4\pi^2t^2}{3T_p^2}}
\end{equation} 
\noindent in which $\beta$ is a factor equal to $1.3801$ that enforces the expression given by equation (\ref{eq:antiSymRicker}) to have a maximum equal to $a_p$.

\noindent The results in Figure (\ref{fig:5}) show a perfect match between analytical results from MATLAB compared to finite element results from OpenSees. Further more, the results under these pulses confirms the use of HHT dissipative time-stepping integration procedure in order to simulate energy dissipation of the rocking motion during the impact.

\subsection{Damped Model}

Considering constitutive law of the viscous damper (\ref{eq:FdVISC}), the damper in the numerical model can be defined as follows. To model the viscous dampers in OpenSees twoNodeLink elements are used \cite{opensees}. This two node link element is placed between two nodes, one at the side of the column and the other one on the ground. The material which is located between this twoNodeLink element is defined with Viscous material available in OpenSees \cite{opensees}. To define this material one need the damping coefficient ($C$) and the power factor ($q$ in equation \ref{eq:FdVISC}, which for the case of linear viscous damper $q=1.0$).

\noindent In this study, to define the damping factor, a viscous damper with similar viscous damping force compared to ones that have been used in South Rangitikei Rail Bridge is used \cite{skinner1980hysteretic}. The dampers in South Ringitikei Rail Bridge are torsianally yielding dampers that has been designed specifically to be used for that bridge \cite{beck1973seismic,skinner1980hysteretic}. The yield force from each torsionally yielding steel beam damper is $F_y=450KN$ \cite{skinner1980hysteretic} therefore, the combined yield force from a pair of dampers is $2F_y=k_d u_y=2\times450kN=900kN$. The equivalent viscous damping force compared to hysteretic damping force can be calculated using the following equation \cite{makris2019effect}:

\begin{equation}\label{eq:ViscVsHyst}
{C_q} = \frac{{{k_d}{u_y}}}{{{{(2b{{\dot \theta }_{\max }})}^q}}}
\end{equation}

\noindent From time history analysis in Figures (\ref{fig:3}) and (\ref{fig:4}) the peak angular velocity ($\dot{\theta}_{max}$) of the responses are ranging from $0.11 rad/s$ to $0.15 rad/s$. According to equation (\ref{eq:ViscVsHyst}), for the pair of torsionally yielding steel-beam dampers with yield capacity $k_d u_y=900kN$, the corresponding damping constant $C_1$ for linear viscous dampers ($q=1$) is $C_1=\frac{900kN}{(2.0m)(0.13rad/sec)}=3461.5 kNsec/m=3461.5 Mg/sec$.
\begin{figure}[b!]
\centering
\includegraphics[scale=0.9]{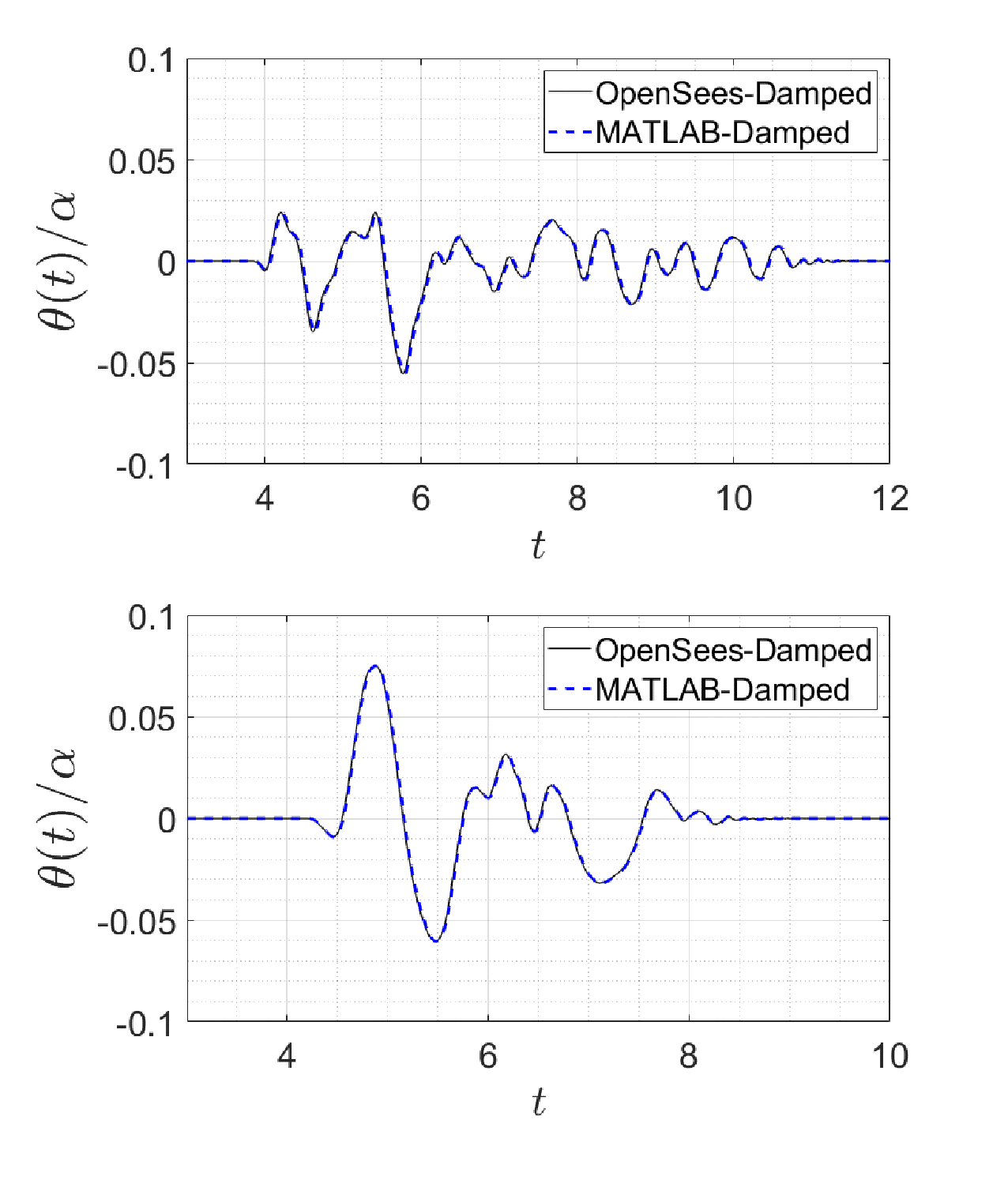}
\caption{Time history response comparison of a rocking column with linear viscous dampers at its pivoting points with damping constant $C_1=3461.5 kNsec/m=3461.5 Mg/sec$ modeled in MATLAB with OpenSees. When subjected to the Newhall/360 ground motion recorded during the 1994 Northridge, California (top) and Takarazuka/000 ground motion recorded during the 1995 Kobe, Japan (bottom) earthquakes.}
\label{fig:6}
\end{figure}

\begin{figure}[t]
\centering
\includegraphics[scale=.9]{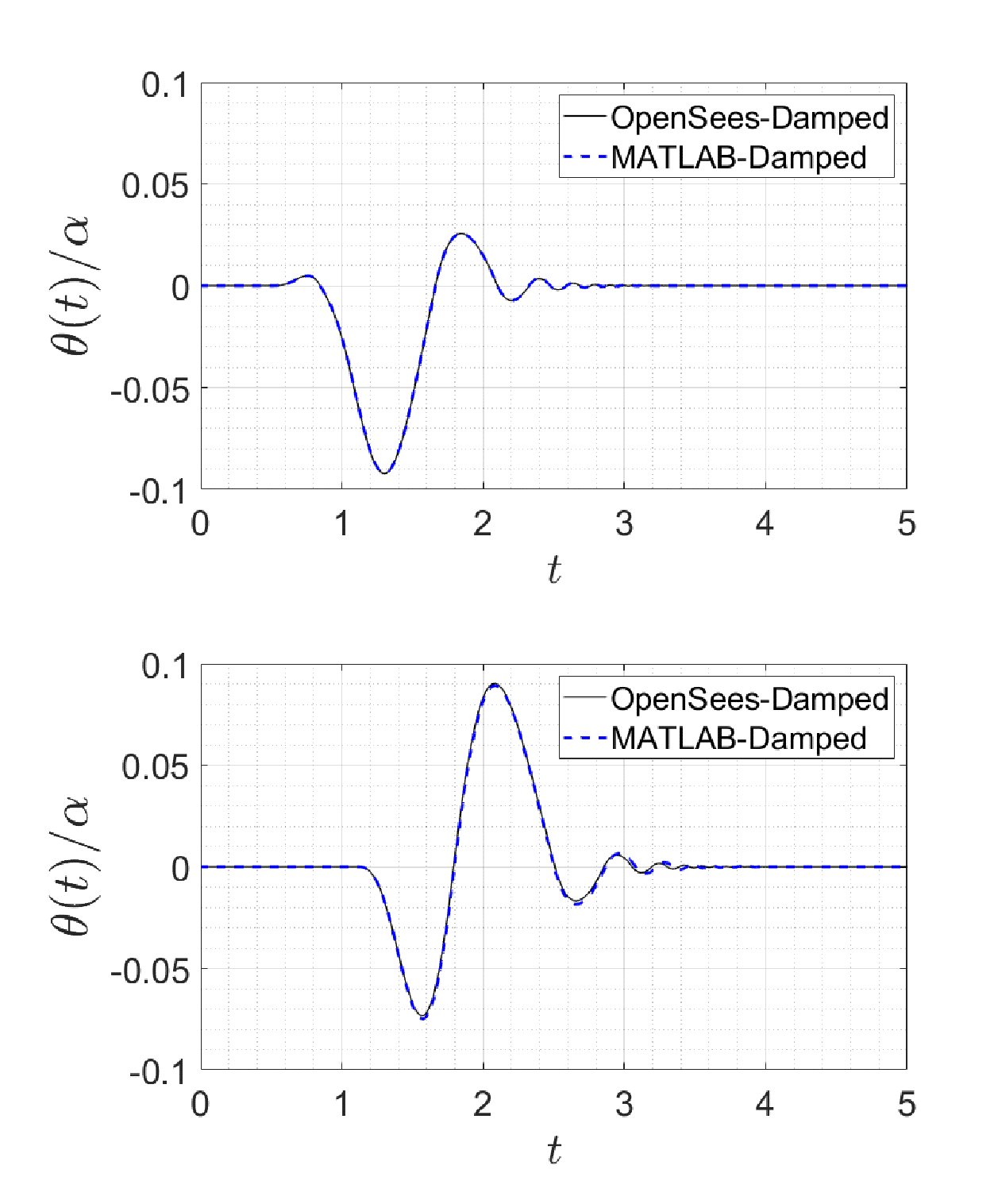}
\caption{Time history response comparison of a rocking column with linear viscous dampers at its pivoting points with damping constant $C_1=3461.5 kNsec/m$ modeled in MATLAB with OpenSees. When subjected to to a symmetric Ricker wavelet (top) and antisymmetric Ricker wavelet with acceleration amplitude ($a_p=0.5g$) (bottom).}
\label{fig:7}
\end{figure}

\noindent Figure (\ref{fig:6}) compares response of the solitary column when the linear viscous dampers with damping constant, $C_1=3461.5 kNsec/m=3461.5 Mg/sec$, are installed at its pivoting points. The top figure is when the system is subjected to the Newhall/360 ground motion recorded during the 1994 Northridge, California and the one in the bottom is when it is subjected to the Takarazuka/000 ground motion recorded during the 1995 Kobe, Japan earthquakes. The results show that use of dampers reduce the maximum rotation of the column and improves it energy dissipation effectively. It is also clear from the results that the finite element model captures the behavior precisely when it is compared to analytical results from MATLAB.

\noindent Figure (\ref{fig:7}) compares response of the solitary column when the linear viscous dampers with damping constant, $C_1=3461.5 kNsec/m=3461.5 Mg/sec$, are installed at its pivoting points. The top figure is when the system is subjected to the symmetric Ricker wavelet and the one in the bottom is when it is subjected to the Antisymmetric Ricker wavelet both with ($a_p=0.5g$). Similar to Figure (\ref{fig:6}), results in Figure (\ref{fig:7}) shows similar trend that the maximum rotation is decreased and analytical MATLAB model supports results from finite element OpenSees model.

\section{Conclusions}
The paper studies finite-element modeling of a solitary rocking column when it is damped at its pivoting points with linear-viscous dampers.

\noindent In first step, the paper derives the nonlinear equation of motion for the rocking column with viscous dampers. Then, the finite-element model for the rocking column is defined in OpenSees using zero-length fiber section with elastic-no-tension material. The energy dissipation of the rocking column when it changes its pivoting point is modeled using Hilber-Hughes-Taylor numerical dissipative time-step integration.

\noindent The finite element model for the solitary undamped rocking column is examined under different earthquake and mathematical wavelet excitation. The results from this stage, confirms the accuracy of the finite element model when it is compared to analytical results from MATLAB.

\noindent The paper also studied the force-displacement relationship of a viscous damper with a hysteretic damper. The results of the force-displacement relation comparison of these two types of damping devices, when they are used along with the rocking block, shows that for the case of the hysteretic damper, the rocking block might experience a greater impact when it changes its pivoting point. This happens because the moment from the hysteretic damping force adopts the same direction as the moment of the self-weight of the block, when the rocking block is rotating back from its maximum rotation to the ground. This phenomenon can cause a greater damage to the rocking-toes and in some cases; it can amplify the responses of the damped block.

\noindent Lastly, the finite element model of the damped system is introduced and the results are matched perfectly with analytical solutions. The study shows that the finite element model introduced, is capable of predicting the behavior of rocking column when it is coupled with dampers with a great accuracy and it can be used in various engineering problems like analyze of damped rocking bridge piers or rocking walls when they are damped.

\setstretch{.8}
\bibliography{references}

\end{document}